\documentclass[twocolumn,letterpaper,10pt]{article}
\usepackage{times}
\usepackage[final]{graphicx}		
\usepackage{citesort}

\begin{document}

\title{\bfseries \Large
SPATIOTEMPORAL DYNAMICS OF PACING IN MODELS OF
ANATOMICAL REENTRY
\vspace*{-12pt}
}
\author{ \large
Sitabhra Sinha ${}^{1,2}$ and David J. Christini ${}^{1,3}$
}
\date{
\vspace*{-8pt}
\normalsize
${}^1$ Division of Cardiology, Weill Medical College of Cornell University, 
New York, NY, 10021, USA\\
${}^2$ Centre for Condensed Matter Theory, Department of Physics, 
Indian Institute of Science, Bangalore 560012, India\\
${}^3$ Dept. of Physiology and Biophysics,  Weill Graduate School of
Medical 
Sciences of Cornell University, New York, NY, 10021\\
}
\maketitle
\thispagestyle{empty}
\normalsize

\emph{Abstract-}\textbf{Reentry around non-conducting 
ventricular scar tissue, which often
causes lethal arrhythmias, is typically treated by rapid stimulation
from an implantable defibrillator. However, the mechanisms of
termination (success and failure) are poorly understood. To elucidate
such mechanisms, we studied pacing of anatomical reentry in 1-D and 2-D
excitable cardiac media models. Our results suggest that
the existence of inhomogeneity in the reentry circuit is essential
for pacing termination of tachycardia to be successful. Considering
the role of such inhomogeneities may lead to more effective pacing
algorithms.\\
\emph{Keywords - } Reentry, tachycardia, antitachycardia, 
pacing}

\begin{center}
\textbf{I. I\textsc{ntroduction}}
\end{center}
Trains of local electrical impulses are widely used to restore
normal wave propagation in the heart during tachycardia - but this
process is not always successful. Such ``antitachycardia pacing'' may inadvertently
induce tachycardias or cause tolerated tachycardias to degenerate to more
rapid and threatening tachyarrhythmias. It is known from clinical
electrophysiology studies that pacing to terminate uniform ventricular
tachycardia (VT) can cause tachycardia acceleration or degeneration to
ventricular fibrillation (VF) \cite{Ros90}. The underlying
mechanisms governing the success or failure of a pacing algorithm are not
yet clear. Understanding these mechanisms is essential, 
as a better knowledge of the processes 
involved in the suppression of VT through
ventricular extrastimuli pacing can aid in the design
of more effective therapies. 
The reentrant action potential may propagate around an inexcitable
obstacle (``anatomical reentry") or may reside in a region of the
myocardium that is excitable in its entirety 
(``functional reentry") \cite{Rud95}.
Although studies
on the role of pacing in eliminating tachycardia have been done for functional
reentry \cite{Dav95}, there has been little work done for the case of
anatomical reentry \cite{Abi95}.

\hspace{.125in} 
Several factors influence the ability of extrastimuli
and/or rapid pacing to interact with VT. 
The most prominent are \cite{Jos93}: (a) VT cycle length (for anatomical
reentry this is determined by the
length of the VT circuit around the obstacle),
(b) the refractory period at the stimulation/pacing site and in
the VT circuit,
(c) the conduction time from the pacing site to the VT circuit, and
(d) the duration of the excitable gap.
As there are a number of conditions to be satisfied before reentry is
successfully terminated, a single extrastimulus is rarely sufficient.
Multiple stimuli are often used - where the first extrastimulus is believed
to ``peel back'' refractoriness to allow the subsequent extrastimuli
to interact with the circuit more prematurely than was possible with only
a single extrastimulus.

\hspace{.125in} 
There has been substantial work on computer modeling of
pacing termination of tachyarrhythmias in a one-dimensional ring 
\cite{Gla95,Rud95}.
The termination of reentry in such a geometry (which is effectively that
of the reentry circuit immediately surrounding an anatomical obstacle)
occurs in the following manner. Each extrastimulus splits into two
branches while traveling around the reentry circuit. The retrograde branch
(proceeding opposite to the direction of the reentrant wave) ultimately
collides with the reentrant excitation, causing mutual annihilation. The
anterograde branch (proceeding along the direction of the reentrant wave)
can, depending on the timing of the stimulation, lead to either
resetting, where the anterograde wave becomes the new reentrant wave,
or termination of reentry, where the anterograde wave gets blocked by the
refractory end of the original reentrant wave. From continuity arguments,
it can be shown that there exists a range of extrastimuli phases 
and amplitudes
that leads to successful reentry termination \cite{Gla95}.
Unfortunately, the argument is essentially applicable only to a 1D model - the
process is crucially dependent on the fact that 
the pacing site is on the reentry
circuit itself. In reality, however, it is unlikely that the pacing site will
be so fortuitously located.

\hspace{.125in}
In this study, we examine the dynamics of pacing from a site some distance
away from the reentry circuit.
One then has to consider the passage of the extrastimulus
from the pacing site to the reentry circuit.
Because the reentrant circuit is the origin of 
rapid outwardly propagating waves,
the extrastimulus will be annihilated before it reaches the circuit under most
circumstances. Successful propagation to the 
reentry circuit will require utilising
multiple extrastimuli to ``peel back'' 
refractory tissue incrementally until one successfully arrives at
the circuit. Further, the anterograde branch must be blocked by the 
refractory tail of the reentrant wave. 
However, as outlined in detail in the Discussion section of our paper,
this is extremely unlikely to happen in a homogeneous medium. As shown by our
simulation results, the existence of inhomogeneity in the reentry circuit seems
essential for pacing termination of VT. 

\begin{center}
\textbf{II. M\textsc{ethodology}}
\end{center}
Modified Fitzhugh-Nagumo type excitable media model equations of
ventricular activation were used to simulate the interaction of
extrastimuli with the VT circuit around an anatomical obstacle. For the
1D model we used the Panfilov model \cite{Pan93,Pan93b} 
defined by the two equations
governing the excitability $e\/$ and recovery $g\/$ variables:
\begin{equation}
\begin{array}{lll}
{\partial e}/{\partial t} & = & {\nabla}^2 e - f(e) - g,\\
{\partial g}/{\partial t} & = & {\epsilon}(e,g) (ke - g).
\end{array}
\end{equation}
The function $f(e)\/$, which specifies fast processes (e.g.,
the initiation of the action potential) is piecewise linear:
$f(e)= C_1 e$, for $e<e_1$, $f(e) = -C_2 e + a$,
for $e_1 \leq e \leq e_2$, and $f(e) = C_3 (e - 1)$, for $e > e_2$.
The function $\epsilon (e,g)$, which determines the dynamics of the
recovery variable, is $\epsilon (e,g) = \epsilon_1$ for
$e < e_2$, $\epsilon (e,g) = \epsilon_2$ for $e > e_2$, and
$\epsilon (e,g) = \epsilon_3$ for $e < e_1$ and $g < g_1$. We
use the physically appropriate parameter values given in Ref.
\cite{Sin01}.
For the 2D studies, in addition to the Panfilov model we used two
other two-variable modified Fitzhugh-Nagumo type models: the
Aliev-Panfilov model \cite{Ali93}, which simulates the restitution
property of cardiac tissue, as well as the model proposed by
Kogan {\em et al} \cite{Kog91}, which
simulates the restitution and dispersion properties of cardiac tissue. 

\hspace{.125in}
We solve the models  by using a forward-Euler integration scheme. We
discretise our system on a grid of points in space with spacing
$\delta x = 0.5\/$ dimensionless units and use the standard three- and
five-point difference stencils for the Laplacians in spatial 
dimensions $d = 1$ and 2 respectively. Our spatial
grid consists of a linear lattice with $L^{\prime}$ points or a square
lattice with $L \times L\/$ points; in our studies we have used values 
of $L^{\prime}$ ranging from 500 to 1000 and $128 \leq L \leq 256$. 
Our time step is $\delta t = 0.022$ dimensionless units. We define
dimensioned time $T$ to be $5$ ms and 
$1$ spatial unit to be $1$ mm. 

\hspace{.125in}
Our initial condition is a wave front at some point in the medium with
transient conduction block on one side to permit wavefront propagation along 
a single direction only. The conduction block is realised by making the
corresponding region refractory. This initiates a re-entrant wave which
roughly corresponds to an anatomically-mediated 
tachycardia. Extrastimuli
are then introduced from a pacing site. In the 1D model, the
ends have periodic boundary conditions - such that the medium represents
a ring of cardiac tissue around an inexcitable obstacle. In the 2D models,
for top and bottom ends we use no-flux (Neumann) boundary conditions since the
ventricles are electrically insulated from the atria (top) and the waves 
converge and annihilate at the region around the apex (bottom). For the sides,
we use periodic boundary conditions.
Two square patches of inexcitable region (whose
conductivity constant is zero) are placed
about the centre of the simulation domain with a narrow passage for
wave propagation between them (Fig. 1). 
No-flux boundary conditions are used at the interface between the
active medium and the inexcitable obstacle.
We use this arrangement to 
represent anatomical obstacles (e.g., scar tissue) around which
``figure-of-eight'' reentry can occur.
This kind of pattern has been seen previously
in actual cardiac tissue (e.g., in post healed myocardial
infarction tissue \cite{Els81}).

\begin{center}
\textbf{III. R\textsc{esults}}
\end{center}
\begin{figure}[t!]
	\centerline{\includegraphics[width=0.85\linewidth,clip] {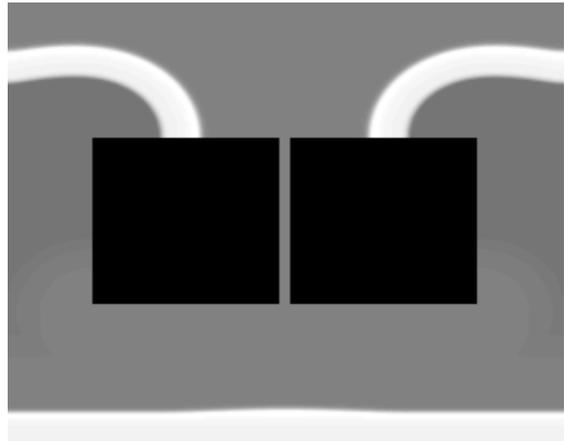}}
	\vspace*{-8pt}

	\caption{\footnotesize Pseudograyscale plot of the $e$-field for
         the 2D Panfilov model ($L$ = 256). 
	 The two square black patches are the
         non-conducting anatomical obstacles. The region of slow 
   	 conduction is in the channel between the two non-conducting
         regions. The figure shows a ``pacing'' planar wave at the bottom of
	 the simulation domain,
         propagating up towards the anatomical obstacles.
	 The earlier pacing-stimulated wave is shown curving 
	 around the obstacles about to enter the narrow channel.}

	\label{fig:fig1}
\end{figure}
In the 2D simulations we looked at the effectiveness 
of different pacing algorithms
in terminating figure-of-eight reentry. 
The pacing was simulated by a planar wave initiated
at the base of the simulation domain. 
This represents a wave that has propagated
a fixed distance up the ventricular wall from a local electrode at the apex -
similar to a wave initiated at the tip of an inverted cone before
propagating up its walls.
Two different pacing patterns were applied. The first used
pacing bursts at a cycle length which was a fixed percentage of the
tachycardia cycle length (determined by the size of the obstacles). 
This corresponds to ``burst pacing'' used in implantable cardiac
defibrillators (ICD) \cite{Hor95}.
The pacing cycle length as well as the number of
extrastimuli in each pacing burst (4-8) were varied
to look at their effects on the reentrant wave. 
The second was decremental ramp pacing
in which the interval between successive extrastimuli was gradually reduced
(corresponding to ``ramp pacing" used in ICDs \cite{Hor95}).

\hspace{0.125in}
For a homogeneous cardiac medium, both ramp and burst pacing were unsuccessful at
terminating reentry (see Discussion for mechanism).
We also looked at the effect of anisotropy by making the
conductivities along the vertical and the horizontal axes 
differ in a ratio of 1:0.3,
which is consistent with 
cardiac tissue \cite{Kee97}. 
As with the homogeneous case,
in the anisotropic case none of the pacing 
methods we tried were successful in 
terminating reentry. 

\hspace{0.125in}
However, we did achieve 
successful termination in the presence of
inhomogeneity in the model. This was done by placing a small zone ( = 7.5 mm) 
of slow conduction
in the narrow channel between the two non-conducting patches. The conductivity in this
region was 0.05 times smaller than the rest of the tissue. A 6-stimulus pacing burst
was found to successfully block the anterograde branch of the extrastimulus traveling
through the narrow channel - and hence terminate the reentry.

\hspace{0.125in}
To understand the process of inhomogeneity-mediated termination we 
implemented a 1D model with a zone of slow
conduction whose length and conductivity constant were
varied to examine their effect on the propagation of the
anterograde branch of the extrastimulus. The reentrant wave was started at a
point in the ring (proximal to the zone of slow conduction) 
chosen to be the origin ($x = 0$) at time $t = T_0 = 0$. At time $t = T_1$,
the first extrastimulus was given at $x = 0$ 
followed by a second extrastimulus at
$t = T_2$ (Fig. 2). The first extrastimulus is able 
to terminate the reentry by itself if it
is applied when the region to one end of it is still 
refractory - leading to unidirectional
propagation. This is identical
to the mechanism studied previously for terminating reentry in a 1D ring
\cite{Gla95}. However, in this study we are interested in the effect of
pacing from a site away from the reentry circuit. 
In that case, it is not possible for the first extrastimulus to arrive
at the reentry circuit exactly at the refractory end of the reentrant wave
(see the Discussion section).
Therefore, we consider
the case when the first extrastimulus is unable to
block the anterograde conduction by itself. 
We have used values of $T_1$ for which
the first extrastimulus can give rise to both 
the anterograde as well as the retrograde
branches.
\begin{figure}[t!]
	\centerline{\includegraphics[width=0.85\linewidth,clip] {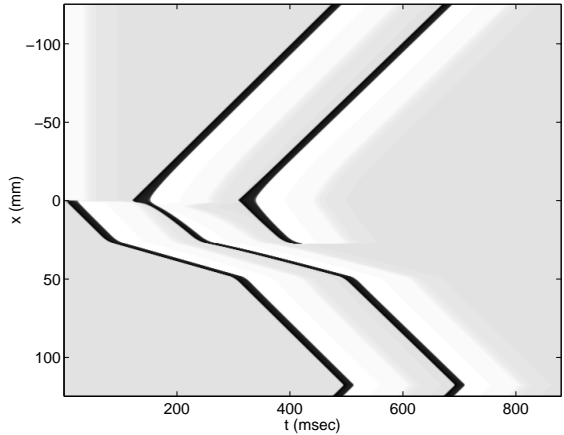}}
	\vspace*{-8pt}

	\caption{\footnotesize Pseudograyscale plot of the e-field 
        showing spatiotemporal propagation
	of reentrant wave in a 1D ring, successfully terminated by pacing
	with two extrastimuli. The zone of slow conduction is between
	$x$ = 23 mm and $x$ = 52 mm. In this region the
	conductivity constant changes from 1 to 0.1 unit with
	a gradient of 0.2 units/mm at the boundaries.
        The reentrant wave is initiated with a stimulus at $x$ = 0 mm
	at $T$ = 0 ms (with transient conduction block in the region
        $x < 0$). The first extrastimulus is
        applied at $T_1$ = 121 ms while the second is applied 
	at $T_2$ = 308 ms (pacing site is $x = 0$ in both cases).}

	\label{fig:fig2}
\end{figure}
Different values of coupling interval ($T_1 - T_0$) and pacing cycle 
($T_2 - T_1$) were used to find which
parameters led to block of the second anterograde wave.
\begin{center}
\textbf{IV. D\textsc{iscussion}}
\end{center}
Because the retrograde branch of the extrastimulus 
always mutually annihilates
the reentrant wave, the success of pacing
in terminating reentry around an anatomical obstacle depends
on whether or not the anterograde branch of 
the extrastimulus is successfully blocked. Although
apparently this is often achieved in practice (i.e., ICDs are often
successful at anti-tachycardia pacing), 
as we have seen in our 2D model simulation, this
is almost impossible to observe  for homogeneous 
tissue unless the pacing site is located on the reentry
circuit. Inhomogeneities appear to be necessary for termination,
as outlined in the following simple mathematical
argument.

\hspace{.125in}
Let us consider a reentrant circuit as a 1D ring of length $L$ with separate
entrance and exit sidebranches (Fig. 3). Further, let the pacing site be located on the
entrance sidebranch at a distance $z$ from the circuit. We use the entrance sidebranch
as the point of spatial origin ($x = 0$) to define the location of the wave on the ring.
The conduction velocity and refractory period at a 
location a distance $x$ away (in the clockwise direction) from
the origin is denoted by $c(x)$ and $r(x)$, respectively. 

\hspace{.125in}
For a homogeneous medium,
$c(x) = c$, $r(x) = r$ ($c, r$ are constants). Therefore, the length of the region in the ring
which is refractory at a given instant is $l = c r$. For sustained reentry to occur, there
must be an area of excitable tissue between the front and the refractory tail of the
reentrant wave  - i.e., the excitable gap. The condition for existence
of this is $L > c ~r$. For convenience, associate $t = 0$ with the time when the reentrant
wave is at $x = 0$ (i.e., the entrance sidebranch) (Fig. 3(a)). 
Let us assume that an extrastimulus
is given at $t = 0$. This extrastimulus will collide 
with the branch of the reentrant
wave propagating out through the entrance sidebranch at $t = z/2c$
(Fig. 3(b)). The pacing site will recover
at $t = r$ and if an extrastimulus is given 
again immediately it will reach the reentry
circuit at $t = r + (z/c)$ (Fig. 3(c)). By this time 
the refractory tail of the reentrant wave will be at
a distance $x = z$ away from the entrance sidebranch and the anterograde branch of the
extrastimulus will not be blocked, i.e., it is impossible for the stimulus
to catch up to the refractory tail in a homogeneous medium. 
This results in resetting of the reentrant wave
rather than its termination.
\begin{figure}[t!]
	\centerline{\includegraphics[width=0.95\linewidth,clip] {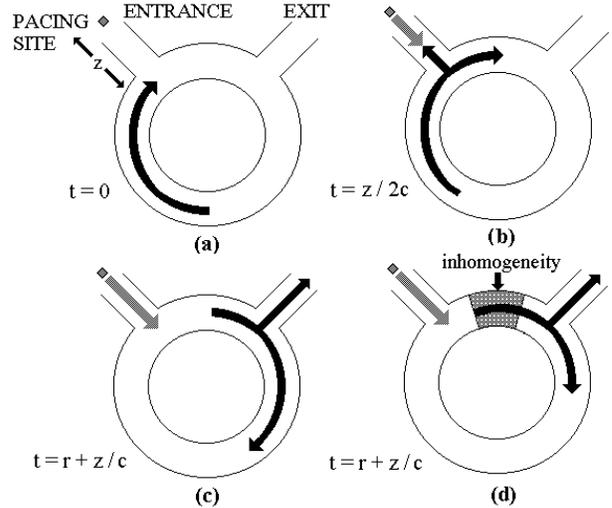}}
	\vspace*{-8pt}
	\caption{\footnotesize Schematic diagram of reentry in 
	a 1D ring illustrating the necessity of a region of inhomogeneity
	for successful termination of reentry by pacing. At $t = 0$
        the reentrant wave reaches the entrance sidebranch (a).
        At $t = z/2c$ the reentrant wave propagating through the
        sidebranch and the first extrastimulus mutually annihilate each
        other (b). At $t = r + (z/c)$ the second extrastimulus
        reaches the reentry circuit by which time the refractory tail is 
        a distance $z$ away from the sidebranch (c). The
        presence of a region of inhomogeneity proximal to the pacing
        site makes it possible that the anterograde branch of the 
        second extrastimulus will encounter a refractory region behind
        the reentrant wave (d).}
	\label{fig:fig4}
\end{figure}
Note that if the first extrastimulus is given at a 
time $t < -z/c$, it reaches the reentry
circuit before the arrival of the reentrant wave. 
As a result, the retrograde branch
collides with the oncoming reentrant wave, 
while the anterograde branch proceeds to
become the reset reentrant wave. It appears that 
only under the very special circumstance
that the reentrant wave reaches the entrance 
sidebranch exactly at the same instant that
the extrastimulus reaches the circuit, is there 
a possibility of the two waves colliding
with each other leading to termination of reentry. 
However, if the circuit has any width,
part of the reentrant wave is likely to 
survive and the wave will continue. Also
such fine-tuning of pacing time is almost 
impossible in reality. As a result,
pacing termination seems all but impossible
if we consider the reentry circuit
to be homogeneous.

\hspace{.125in}
The situation changes, however, if an inhomogeneity 
(e.g., a zone of slow conduction) exists in the
circuit (Fig. 3(d)). In this case,
the above argument no longer holds as the waves travel 
faster or slower depending on their
location in the circuit. If the pacing site is proximal to the zone of slow conduction,
then the pacing site may recover faster than points on the reentry circuit. As a result,
extrastimuli may arrive at the circuit from the pacing site and encounter a region that
is still recovering. This leads to successful block of 
the extrastimulus anterograde wave, while
the retrograde wave annihilates the reentrant wave, 
resulting in successful termination.
The success of pacing will depend on the location 
of the inhomogeneity. If pacing is
performed distal to the zone of slow conduction, 
termination will be harder to achieve
as the extrastimulus will have a longer distance 
to traverse before reaching the
inhomogeneity, which will correspondingly 
have a longer time in which to 
recover. This is supported
by electrophysiological studies of pacing in cardiac tissue \cite{Ing00}.

\hspace{.125in}
Similar arguments may apply for other types of inhomogeneity. 
For example, existence
of a region having longer refractory period than normal tissue will lead to the development
of patches of refractory zones in the wake of the reentrant wave. 
If the anterograde branch of the extrastimulus arrives at such a zone before 
it has fully recovered, it will be blocked. In fact, a computer modeling study
of the interruption of tachycardia through pacing 
has been performed assuming the existence of
such a region on the reentrant circuit \cite{Abi95}.

\hspace{.125in}
By providing simulation results of 1D and 2D 
models in which pacing is unsuccessful
in terminating anatomical reentry in the 
absence of any inhomogeneities and providing
a simple mathematical argument in support of this conclusion, this paper
outlines the potential significance of inhomogeneities 
in cardiac tissue in the
termination of VT through pacing.
\vspace{2pt}

\begin{center}
\textbf{V. C\textsc{onclusion}}
\end{center}
There are obvious limitations of our study - the most prominent being the
use of excitable media models of ventricular activity. These models do not
incorporate details of ionic currents (unlike, e.g., the Luo-Rudy
model) but rather lump such details into a single model parameter.
Further, we have used 1D or 2D models in our
study, even though the heart's 3D structure is important in reentry. 
We have also assumed
the heart to be a monodomain rather than a bidomain (which has
separate equations for intracellular and extracellular space) - for the 
low antitachycardia pacing stimulus amplitude
we believe this
simplification to be justified.

\hspace{.125in}
Despite these limitations, the results presented here may apply to the
case of pacing termination of anatomical reentry in the ventricle.
We have tried to develop general (i.e., model-independent) mathematical
arguments about the requirement that inhomogeneities should exist
in the reentrant circuit for pacing to be successful in eliminating VT.
\begin{center}
\textbf{A\textsc{cknowledgment}}
\end{center}
We thank Ken Stein for helpful discussions.
This work was supported by the American Heart 
Association (\#0030028N).

\vspace{2pt}

\begin{center}
\textbf{R\textsc{eferences}}
\end{center}

\vspace*{-8pt}

\end{document}